\documentclass[conference]{IEEEtran}
\IEEEoverridecommandlockouts
\usepackage{cite}
\usepackage{amsmath,amssymb,amsfonts}
\usepackage{algorithmic}
\usepackage{graphicx}
\usepackage{textcomp}
\usepackage{xcolor}

\def\BibTeX{{\rm B\kern-.05em{\sc i\kern-.025em b}\kern-.08em
    T\kern-.1667em\lower.7ex\hbox{E}\kern-.125emX}}
    
\usepackage{booktabs} 

\newcommand\todo[1]{\textcolor{red}{#1}}
\newlength{\hcolw}
\newcommand\scale[2]{\begin{minipage}{#1\textwidth}\centering{#2}\end{minipage}}

\graphicspath{{./media/}}

\begin{document}

\title{Analyzing HPC Support Tickets:\\Experience and Recommendations
}

\author{\IEEEauthorblockN{Alexandra DeLucia$^*$\thanks{$^*$Work performed during an internship with the Ultrascale Systems Research Center at Los Alamos National Laboratory.}}
\IEEEauthorblockA{Department of Computer Science\\
Johns Hopkins University\\
Baltimore, MD \\
aadelucia@jhu.edu}
\and
\IEEEauthorblockN{Elisabeth Moore}
\IEEEauthorblockA{Information Sciences Group\\
Los Alamos National Laboratory$^{\dagger}$\thanks{$^{\dagger}$This manuscript has been approved for unlimited release and has been assigned \todo{LA-UR-19-28780}. This work has been authored by an employee of Triad National Security, LLC which operates Los Alamos National Laboratory under Contract No. 89233218CNA000001 with the U.S. Department of Energy/National Nuclear Security Administration.
	The publisher, by accepting the article for publication, acknowledges that the United States Government retains a non-exclusive, paid-up, irrevocable, world-wide license to publish or reproduce the published form of the manuscript, or allow others to do so, for United States Government purposes.}\\
Los Alamos, New Mexico \\
lissa@lanl.gov}
}

\maketitle

\begin{abstract}
High performance computing (HPC) user support teams are the first line of defense against large-scale problems, as they are often the first to learn of problems reported by users. Developing tools to better assist support teams in solving user problems and tracking issue trends is critical for maintaining system health. Our work examines the Los Alamos National Laboratory HPC Consult Team's user support ticketing system and develops proof of concept tools to automate tasks such as category assignment and similar ticket recommendation. We also generate new categories for reporting and discuss ideas to improve future ticketing systems. 

\end{abstract}


\section{Introduction}\label{sec:intro}
High performance computing (HPC) user support teams are the first line of defense against large-scale problems, as they are often the first to learn of problems reported by users. These teams are the first to investigate system and software issues, and they occasionally assist users with installing and building their applications as well as diagnosing application failures. They also are in charge of tracking problems, notifying users of system downtime and maintenance, and maintaining internal guides and FAQs. The user support team is crucial for running large-scale computing facilities, but often experiences pain points even with access to currently available analysis and tracking tools. 

There are many request tracking (\textit{ticketing}) systems available, but very little published research about them. As discussed in Section \ref{sec:related}, most of the research is done in-house or only by the ticketing software companies. We believe open research is important in this field, especially researching ticketing systems for tracking HPC issues. Even though ticketing systems have similar workflow across industries, HPC-specific research would help with integrated system design to combine information across clusters, filesystems, and users, and for developing tools for faster problem solving. 

In this work we collaborated with the Los Alamos National Laboratory HPC user support team, known locally as the Consult Team, members of which are referred to as ``consultants''. They provided us with historical data from their ticketing system and general guidance regarding current pain points. Throughout the process, we kept track and contributed to those ideas, which are presented in Section \ref{sec:improvements}.

In addition to ticketing system suggestions, we created proof of concept tools to automate aspects of the ticketing process, specifically automatically assigning a category to a ticket (e.g. ``Environment-System", ``Accounts"), and recommending similar tickets, detailed in Sections \ref{sec:category_prediction} and \ref{sec:ticket_recommend}, respectively. Ultimately, we would like to create a similar ticket recommendation system, and requirements for this are discussed in Sections \ref{sec:ticket_tools} and \ref{sec:improvements}.

As tools for automation are important for streamlining the ticketing workflow, so are methods for analyzing trends in community interactions and techniques for general text analysis of the ticket content. ``Community" refers to groupings of consultants, users, and machines. This is discussed in Section \ref{sec:exploration}. In our text analysis, we focus on clustering words to assist with defining categories for reporting.

Our work contributes the following:
\begin{itemize}
	\item Proof of concept tools to automate ticket category assignment and similar ticket recommendation
	\item Discussion of experience with ticketing community analysis for discovering trends
	\item Machine learning-based text analysis of ticket content for exploration and category generation
	\item Suggested improvements to the HPC ticketing systems to make analytics and add-on tool development easier
\end{itemize}

\section{Related Work}\label{sec:related}
Due to the use of request tracking applications across various industries, it is surprising that there is little published work on analysis of these systems. This is probably due to most research happening ``in-house" or only within companies that specialize in proprietary subscription-based applications such as ZenDesk\footnote{https://www.zendesk.com/} and ServiceNow\footnote{https://www.servicenow.com/}. 

Some companies have published information on their internal workflow, such as Uber \cite{DBLP:journals/corr/abs-1807-01337} and IBM \cite{DBLP:journals/corr/abs-1711-02012, Zhang2010PTMPT}. Published work focuses on automatic solutions and problem categorization. Mani \textit{et al.~}create an automated system that matches a user's question to an answer by classifying the question into a ``question type" and then using features from the question to match it to a solution in a large internal Apache Lucene\footnote{https://lucene.apache.org/. Elasticsearch is based on Lucene} knowledge base of all related website pages and documents (e.g. PowerPoints, PDFs) \cite{DBLP:journals/corr/abs-1711-02012} . This is an impressive system, however it could only be implemented with a dataset of many quality question/solution pairs. Creating this dataset would require manually labelling thousands of tickets, or be created over time by recording a ticket solution in the system. Zhang \textit{et al.~}also use a set of QA pairs to train a system using probabilistic topic mapping to match questions with solutions \cite{Zhang2010PTMPT}. Sometimes a ``solution" is a standard template, as in work by Molino \textit{et al.~}, who compare a topic modeling-based random forest model to a deep learning approach for matching a request to a reply template \cite{DBLP:journals/corr/abs-1807-01337}. Our work does not look into question/solution matching due to lack of a dataset, discussed in Section \ref{sec:improvements}. Instead, we use unsupervised approaches by necessity. Published work also focuses on automatic categorization of user problems, like the categorization of a question into a question type in \cite{DBLP:journals/corr/abs-1711-02012}, and the categorization of a question into a ``contact type'' as in \cite{DBLP:journals/corr/abs-1807-01337}. We create a model for automatic categorization of tickets into reporting categories (e.g. ``Environment-System").

\section{Available Data}\label{sec:avail_data}
In this work we analyze tickets from the Los Alamos National Laboratory High Performance Computing (LANL HPC) Consult Team. The team sits in the Environments group within the HPC division, and is responsible for helping users by troubleshooting jobs, answering questions, and assisting other HPC groups by investigating system issues. The team is also responsible for creating and maintaining online user documentation for the LANL HPC user community (e.g. FAQs for installing software, submitting jobs, etc.). The users, referred to within the ticketing system as \textit{requestors}, are lab scientists or their industry and academic collaborators.

\begin{figure}
	\centering
	\includegraphics[width=\columnwidth]{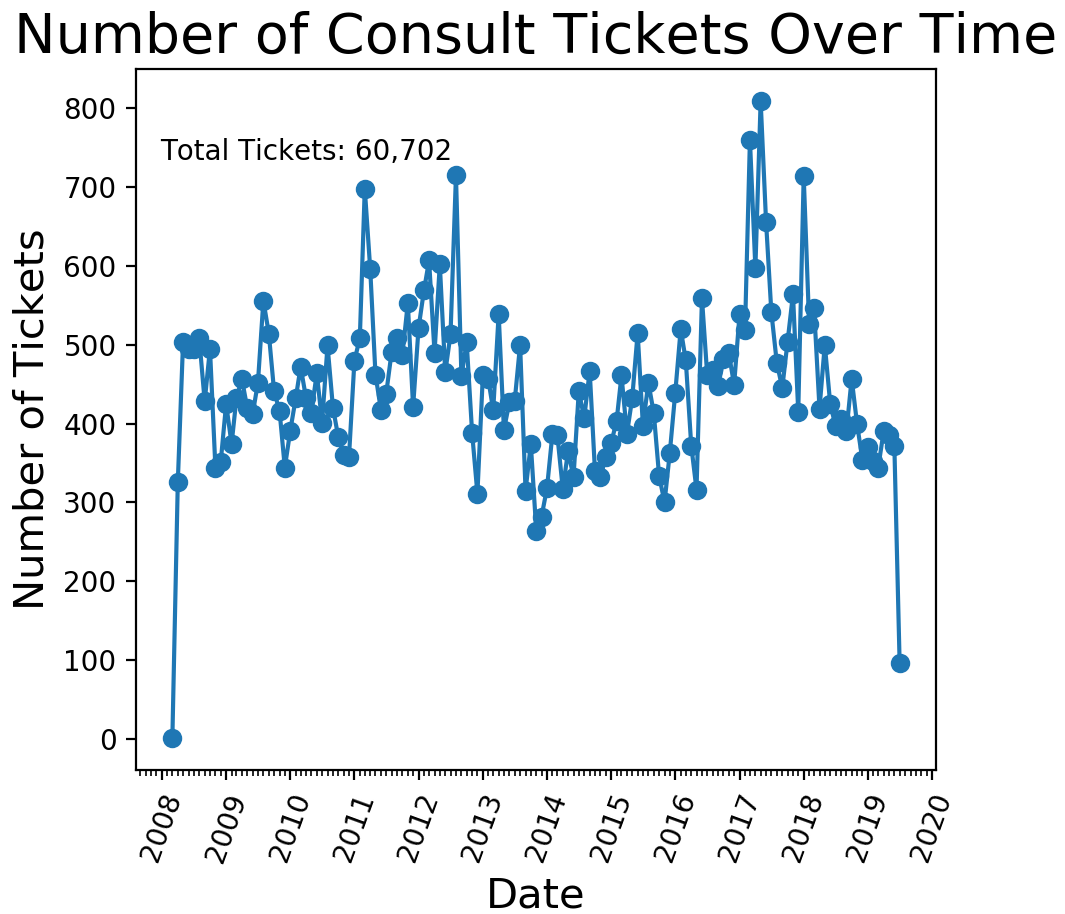}
	\caption{Plot of the LANL HPC Consult team Request Tracker (RT) ticket dataset over time by month. There were a total of 60,702 tickets between 2008 and 2019. These numbers do not include rejected or deleted tickets.}
	\label{fig:tickets_over_time}
\end{figure}

The Consult Team uses Request Tracker (RT) 4.2.10 by Best Practical\cite{rt} to track user problems. This is an open source, customizable system that works via email communication with the requestor and has an owner-facing (i.e. consultant) interface to assign categories, add correspondences, and track tickets. The team has been using the system since 2008. Other groups within the HPC division also use the system, but the Consult Team has its own ``queue." We only examine tickets that belong to this Consult queue. For all tickets we gathered the ticket metadata (date created, assigned categories, user, consultant) and content (emailed correspondences, consultant comments). We divided the ticket content into consultant comments, ticket create message (the initial contact message for emailed tickets), and the full ticket content (all emailed correspondences, issues with this full content is discussed later). The fields available for each ticket are shown in Table 
ref{\label{tab:request_anatomy}}. Our dataset contains a total of 60,702 tickets after the removal of rejected (invalid) tickets (shown in Figure \ref{fig:tickets_over_time}).

\subsection{Accessing and Storing the Data}
We used a modified Python wrapper for the RT REST API to query the tickets \cite{python-rt}. The REST API provided a simpler interface for querying all ticket metadata and content than querying the RT MySQL backend directly, however it does not access deleted tickets. Gathering the email correspondences was also simpler with this API. However, due to the communication through email, many of the ticket transactions have reply chains, which leads to repeated information in each transaction. Not all users continue to reply with the entire quoted message, so using only final transaction was not an option. For the full content dataset, we left the emails with the repeated correspondences. The RT front-end seems to know where the quoted message starts, so an email parser could possibly fix this.

We stored the data locally in text files. Since there were under 70,000 tickets, this was feasible and faster than performing an API request. We attempted to primarily store the data in  Elasticsearch\footnote{https://www.elastic.co/}, but decided against it as we needed access to the text tokenization. Elasticsearch, essentially a wrapper for Apache Lucene, provides customization for how text should be tokenized and which similarity metric to use, but did not provide an easy way to request the vectors or tokens for all the tickets in an easy pipeline for feature engineering. Further, when we tried making a dashboard with Kibana, the interface was more difficult than making a custom one with something like Dash.\footnote{https://dash.plot.ly/}

\subsection{Cleaning the Data}\label{sec:data-cleaning}
We do not alter the ticket metadata, but the ticket content is cleaned using standard natural language processing techniques (NLP). We process the content as follows:
\begin{enumerate}
	\item remove non-English characters
	\item replace the automated reply footer with `` footer "
	\item convert the content to lowercase
	\item remove domain-specific stop words (tickets, ticketing, mailto, wrote, re, fwd)
	\item replace phone numbers with `` phone\_number "
	\item preserve email addresses and URL/file paths by replacing ``http(s)?://" with ``http\_" and ``@" and ``-" with ``\_"
	\item remove general symbols
	\item replace hex with `` hex\_number "
	\item stem words with NLTK Snowball Stemmer \cite{Loper02nltk:the}
	\item join defined bigrams (high performance computing, los alamos, etc.)
	\item replace all whitespace with a single space
\end{enumerate}
The regular expressions for the cleaning and tokenization are in Table \ref{tab:cleaning_regex}. For greater flexibility, we did not store the ``clean" data with removed English stop words or with pre-defined tokens. This was so that we could experiment with the types of tokens and inclusion of stop words (e.g. whether to include file paths or tokens with only alphabetical characters).

\begin{table*}
\centering
\caption{Regular expressions used in cleaning and tokenizing the ticket content for analysis. The regex was implemented in Python and might need to be modified for other languages. Note: the string patterns matching the beginning and end of strings and whitespace is used instead of the default word boundaries to avoid breaking on non-alphanumeric characters (keeps URLs intact).}
\begin{tabular}{cc}\toprule
\textbf{Task} & \textbf{Regex} \\ \midrule
non-English characters & \verb|r"[^\x00-\x7f]"| \\ \midrule
phone numbers & \verb~"(\d{3}-\d{3}-\d{4})|\d{3} \d{3}-\d{4}"~\\ \midrule
hexadecimal & \verb~"0x[0-9a-f]+|[0-9a-f]{16}"~ \\ \midrule
unwanted symbols & \verb@"[!#<>:\[\]\{\}€,\"\(\)\*;]+|[\.\?]\s|:[-_~=\.]{2,}"@\\ \midrule
Token pattern: only alphabetical characters & \verb~r"(?:\s|^)([a-z]{2,})(?:\s|$)"~ \\ \midrule
Token pattern: alphanumerical that start with a letter & \verb~r"(?:\s|^)([a-z]\w+)(?:\s|$)"~ \\ \midrule
Token pattern: alphanumerical and include URLs and file paths & \verb~r"\b[a-zA-Z\/][\w\/\?\.\=]+\b"~ \\ \bottomrule
\end{tabular}
\label{tab:cleaning_regex}
\end{table*}

\begin{table*}
\centering
\caption{Basic anatomy of a ticket. There are other fields available in the RT system, but they are not used (e.g. request priority).}
\begin{tabular}{cc}
\toprule
\textbf{Component} & \textbf{Description} \\ \midrule
Requestor & Person who submitted the request (i.e. user)\\ \midrule
Consultant & Employee assigned (or self-assigned) to resolve requestor's issue\\ \midrule
Date & Creation date of ticket\\ \midrule
Contact Method & How the requestor contacts customer service (i.e. email, phone)\\ \midrule
Status & \begin{minipage}{\hcolw}Current status of ticket. (i.e. deleted, new, open, rejected, resolved, stalled, waiting on user)\end{minipage}\\ \midrule
Subject & \begin{minipage}{\hcolw}Brief subject line for request. It is user-generated when the request is created via email, and can be Owner-generated when the request is submitted over the phone.\end{minipage}\\ \midrule
Category & \begin{minipage}{\hcolw}Label describing the nature of the request. In the RT system, a ticket can be assigned multiple categories but usually only one is used.\end{minipage}\\ \midrule
Comment & \begin{minipage}{\hcolw}Owner notes that are not sent to the user. Usually to track attempted solutions for future similar problems.\end{minipage}\\ \midrule
Article & \begin{minipage}{\hcolw}A ``reply template" a consultant can send to a Requestor (e.g. "submit form A to so-and-so to open ports on machine X")\end{minipage}\\ \midrule
Transaction & \begin{minipage}{\hcolw}History of updates to the request. Includes free-text correspondences and assigned metadata (i.e. updating status and assigning a category).\end{minipage}\\ \bottomrule
\end{tabular}
\label{tab:request_anatomy}
\end{table*}

\section{Ticketing Tools}\label{sec:ticket_tools}
\subsection{Automatic Assignment of Ticket Category}\label{sec:category_prediction}
\begin{figure}
	\includegraphics[width=\columnwidth]{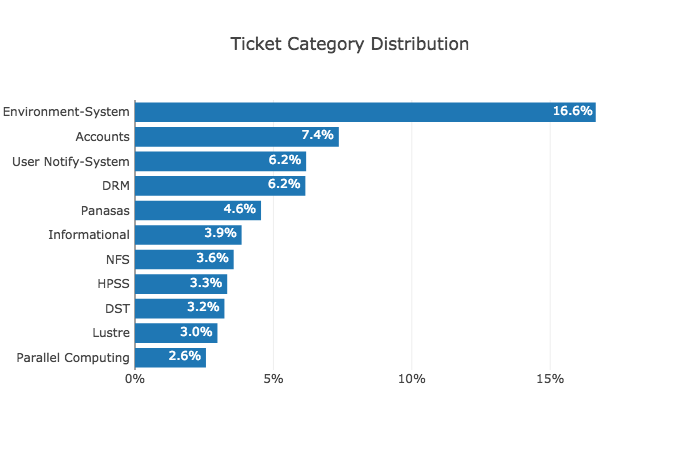}
	\caption{Distribution of ticket categories. There are 93 total categories but the above categories were assigned to at least 2\% of tickets.}
	\label{fig:category_distribution}
\end{figure}

Correct assignment of a ticket to a descriptive category is important because it is used for reporting and tracking the types of issues the consultants have encountered. Choosing an appropriate category can be time consuming, and occasionally, subjective to the consultant. We developed a model that can suggest a category for a ticket based on the incoming create message and subject. This suggested category could help save time and assist with ticket triaging and reduce some category subjectivity by providing options. The Consult Team has 93 pre-defined categories, but only a handful are frequently selected. A breakdown of ticket category is in Figure \ref{fig:category_distribution}.

\subsubsection{Methods}
We created a dataset from resolved tickets that were emailed (so that we had access to the initial create message from the user) and were in a currently used category that had at least 10 tickets. We discarded the multi-categorized tickets (2\% of all tickets) for ease of model evaluation. With this dataset of 23,385 create messages and subjects labeled by assigned category, we used standard text analysis techniques to create a numerical representation of the text for a machine learning classifier to predict the assigned category. We created vector representations (``feature sets") of the cleaned ticket subject lines and create messages with latent Dirichlet allocation (LDA) topic modeling\cite{McCallumMALLET, Blei:2003:LDA:944919.944937}, latent semantic analysis (LSA) \cite{scikit-learn}, and Doc2Vec\cite{rehurek_lrec, DBLP:journals/corr/LeM14}. The algorithm and model parameters are in Table \ref{tab:predict_cat_settings}. We included all alphanumerical tokens, URLs, and file paths (see Table \ref{tab:cleaning_regex}). We used a Scikit-learn implementation of random forest classifier\cite{scikit-learn} to predict the assigned ticket category. The models were evaluated using weighted precision-recall metrics and ``accuracy at 3." Precision is the ratio of tickets that the model correctly predicted as category X to all tickets predicted as category X, recall is the ratio of tickets that were predicted as category X to all tickets assigned category X, and F1 is the harmonic mean of the two. For a precision-recall score, a value closer to 1 is better. Accuracy at 3 means a model's prediction is correct if the assigned category is in the top 3 predicted classes. This metric is good for evaluating how useful the model would be as a category suggestion tool. For this experiment, we considered the assigned category to be the ``correct" category, and did not ask other consultants for second opinions. We discuss the subjectivity and complications of category assignment in Section \ref{sec:improvements}. 

\begin{table}
\centering
\caption{Parameters for generated feature sets.}
\begin{tabular}{cc}\toprule
LDA 10 topics &\scale{0.3}{topics=10, iterations=1000, alpha=10, beta=0.005, stopwords removed} \\ \midrule
LDA 500 topics & \scale{0.3}{topics=500, iterations=1000, alpha=0.01, beta=0.005, stopwords removed} \\ \midrule
LSA & dimension=100, stopwords removed \\ \midrule
Doc2Vec & \scale{0.3}{min count=5, dimension=400, context/window=10 (Note: we used the normalized vectors)} \\ \bottomrule
\end{tabular}
\label{tab:predict_cat_settings}
\end{table}

\begin{figure}
	\includegraphics[width=\columnwidth]{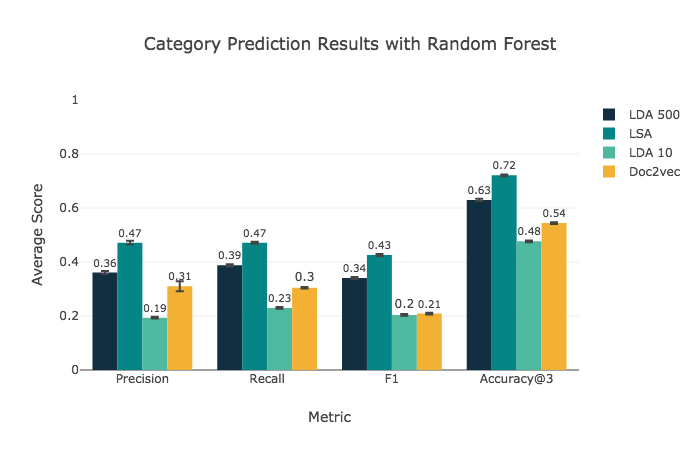}
	\caption{Results from predicting assigned ticket category using each feature set with a Random Forest model (10 estimators). We averaged 200 trials for the results.}
	\label{fig:category_prediction}
\end{figure}

\subsubsection{Results}
When evaluated on prediction, the model achieved F1 scores of 20\%-40\%, and performed considerably better when evaluated on suggestion (i.e. ``accuracy@3"), with scores of 50\%-70\%. The LDA 500 and LSA feature sets performed the best across all metrics. We think the generally low model performance can be attributed to the vast number of categories (we explore new categories in Section \ref{sec:cat_anal}). We also considered the idea that not all problems are readily apparent from just the user's initial message, but roughly 60\% of all resolved emailed tickets were assigned a category by a human after viewing only the create message. For future work, model performance could be improved by including different tokens (see Table \ref{tab:cleaning_regex}) or tweaking classifier parameters. The results are shown in Figure \ref{fig:category_prediction}.

\subsection{Ticket Recommendation}\label{sec:ticket_recommend}
When a consultant begins a ticket, the only way to see if there are similar problems is to ask the other consultants or attempt a search based on the details they remember (requestor, time frame, specific wording). The Consult RT system has a full context search feature but it is not very effective or efficient, as evidenced by consultants opting to search only the metadata fields. The ability to see tickets similar to any current tickets would help a consultant better and more quickly assist a user by viewing previous solutions to similar problems. A ticketing system with a tool that recommends similar tickets based on the create message of an incoming ticket would be useful for not only solving one user's problem, but for tracking problems over time. If a ticket is similar to tickets that have been appearing throughout the day, then there might be a system-wide problem. However, if similar tickets come in intermittently over months, then the solution would be useful to make into a template or add to the internal resources website, like a FAQ. The main use cases of researching similar tickets are as follows:
\begin{itemize}
	\item Solving a current ticket by looking at previous or related completed tickets
	\item Reporting the occurrences of specific issues
\end{itemize}

\begin{figure*}
	\centering
	\includegraphics[width=.75\textwidth]{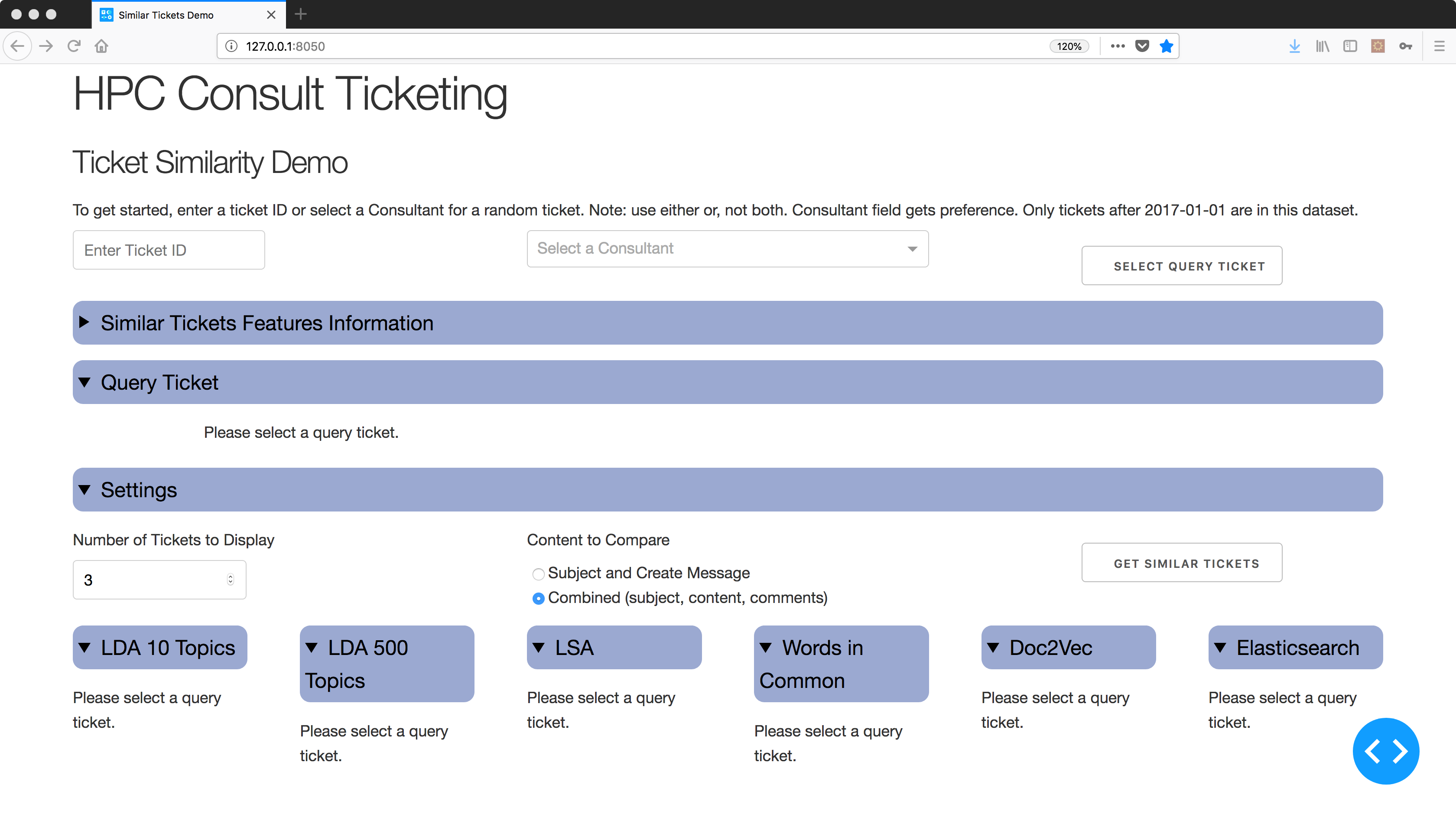}
	\caption{Screenshot of the similarity demo presented to consultants for ticket similarity model evaluation.}
	\label{fig:similarity_demo}
\end{figure*}

\subsubsection{Methods}
For this similarity demo, we used a subset of the tickets from January 2017 to July 2019. We applied the same algorithms and parameters as the category prediction (Section \ref{sec:category_prediction}), shown in Table \ref{tab:predict_cat_settings}, to the create message and subjects and to the ``combined" ticket content (i.e. subject, create message, content, and comments). The feature sets were re-generated on the subset of data. As a pseudo-baseline we included the results from an Elasticsearch ``more like this" query and a ``naive" calculation of percent words in common. For the LDA, LSA, and Doc2Vec feature sets, we say a ticket is ``similar" to another ticket if their representative vectors have a cosine similarity close to 1, where cosine similarity = 1 - cosine distance. 

\subsubsection{Issues with Model Evaluation}
For feature set comparison, we created a simple GUI that takes a ticket ID and shows the top 3 results from each feature set, shown in Figure \ref{fig:similarity_demo}. Even though each feature set returns a score of how similar a ticket is to another ticket, this score represents the closeness of the ticket vector representations (cosine similarity), not necessarily the return tickets' helpfulness or relatedness. We need human experts for that evaluation. We also ran an experiment with a random sample of 200 tickets and collected information on the ticket category. Because of this, our presented results are a mix of statistics of model results and qualitative review of results based on a small informal discussion with LANL HPC Consult Team. For the discussion, we presented the demo with the ``combined" ticket content and primarily discussed the query results of a very common ticket, a consultant-created weekly email draft for general user information, and a user-created ticket concerning a specific Firefox issue. 

\subsubsection{Results}
\paragraph{Model Results}
The results from our experiment supported our notions on the relatedness of tickets in a single category.  From the trials with 200 sample tickets, on average under 1 ticket in the top 3 returned from the feature sets shared a category with the query ticket. This follows the results that category is not a good metric for ticket relatedness. Further, it was pointed out to us by a consultant that only returning tickets in the same category is not especially useful since they already have the ability to query by category. Because of this, our presented results are purely qualitative and based on a small informal discussion with consultants.

\paragraph{Discussion}
Since the similar ticket demonstration only presented a couple of tickets and their query results, there was not enough information to determine which feature set model (e.g. Doc2Vec) overall performed the ``best." However, there was a similarity in the kinds of results. Between LDA 10 and LDA 500, LDA 500 would pick up on more specific keywords, which makes sense due to the higher granularity of 500 topics instead of only 10. This is a similar pattern to the LDA feature sets performance on the category assignment task from Section \ref{sec:category_prediction}. 

On the example common ticket, a weekly email draft, all feature sets performed very well by returning other weekly email drafts. This set an easy baseline. On the next task of picking up on information from a more specific ticket, the results were mixed. Elasticsearch did the best on returning tickets that were most similar to the original ticket, and the others tended to have more of a variety in their results, where one ticket would be very similar and the others related in some way. We were told that both types of results could be helpful for the main use cases mentioned earlier. Looking at more of variety of related tickets could be more beneficial for problem solving, but finding tickets with the exact same issue is better for reporting and tracking issues. A possible solution is to use different feature set models for different tasks, such as the more specific Elasticsearch model for querying tickets for reporting and then LDA 500 for reading more general, but still related, tickets for researching solutions.

A concern that came up was the models' ability to pick up on variations in vocabulary for the same concept, beside misspellings and verb tense which are dealt with stemming, such as ``machines" and ``computers", or ``permissions" and ``policies." With this concern we expect Doc2Vec to fare better because of how it represents words based on their surrounding context, which would be similar across vocabulary for the same concept. We show the results of a Word2Vec word similarity model, which is part of Doc2Vec, in Table \ref{tab:word2vec_query}. As for general UI feedback, we were asked for the option to show more tickets (default was to show top 3 closest tickets) and to filter (or weight) the results by metadata such as consultant or date range.

\subsection{Automatic Ticket Reply}\label{sec:auto_reply}
The LANL HPC Consult Team has a set of templates created for standard procedures, such as requesting a port to be opened. In the RT system, the template is called an ``article" (see Table \ref{tab:request_anatomy}) and can be inserted into the email reply of a message. The consultant can alter the template as needed and then send it to the user. Similar to COTA \cite{DBLP:journals/corr/abs-1807-01337}, we tried to predict which template to send, if any, to an incoming ticket. 

The way that RT handles articles was not clear, and we were told much later by a consultant that when an article is used it is recorded as a ``Refers to" ticket link. Before that knowledge, we created an Elasticsearch instance and filled it with the ticket content and had copies of text of the 22 templates consultants can access from their queue. We used the basic search feature to query the tickets using the body of each article and then manually checked the query results. With this method only 10 of those templates were found in a single ticket and some of those templates were too similar to immediately differentiate (e.g. a notice of a ``runaway" job versus a final notice of a runaway job and requesting access to x versus requesting access to y). We did not find enough tickets with templates to create a dataset for model training and testing. After we were told that RT does record article use, we checked and realized our premise that consultants utilized the template feature was incorrect. We discuss the idea of templates more in Section \ref{sec:improvements}.

If we did have a dataset, we would have used the same features as the automatic category assignment (see Section \ref{sec:category_prediction}), with the template as the prediction target.

\section{Ticket Exploration and Analytics}\label{sec:exploration}
This section reviews miscellaneous tools and ideas we wanted to apply to the dataset for the purpose of uncovering patterns. We look into analyzing the consultant and user interactions and revisit the concept of categories. We specifically stayed away from looking at anything resembling a performance metric (i.e. consultant response time, ticket closure rate). These are already known evaluation methods and outside our scope of getting insights on problems and system use and developing tools that would help consultants and streamline the ticketing process.

\subsection{Community Analysis}
We represented the ticketing ``community" as a graph by assigning nodes to the consultants and users and weighting edges by number of tickets. The graph is not shown here because it was too large to visualize usefully and did not reveal any interesting macro-level trends.  Restricting the graph to only a single user or consultant and including nodes for Category and machines could uncover trends in user and machine issues. Another way to reduce the scope of the graph would be to group users by project or lab division. Unfortunately we did not have access to that information, since the ticketing system is separate from other employee information systems. Ideas for adding outside information to the ticketing system is in Section \ref{sec:improvements}.

\subsubsection{A Note on Named Entity Recognition (NER)} Since we were looking at the community, we thought it would be interesting to see what a named entity recognition tool would find for us. We used the StanfordNER tool \cite{Finkel:2005:INI:1219840.1219885}. Unfortunately, the results were not precise enough to be useful, however the tool did pick up on cluster names and identified them as proper nouns. We realized that using the metadata fields, or even a known list of cluster and program names, to search for in the tickets would give us better results than an NER tool. An interesting application would be training an NER classifier to find new software that users ask questions about, as maintaining a list of all new software would be tedious.

\subsection{Category Analysis}\label{sec:cat_anal}
As discussed in Section \ref{sec:category_prediction}, the ticket metadata field of categories are used for reporting. From our discussions with the Consult Team, we were intrigued by the trends in category usage and defining categories.

\subsubsection{Category Trends}
\begin{figure}
	\centering
	\includegraphics[width=\columnwidth]{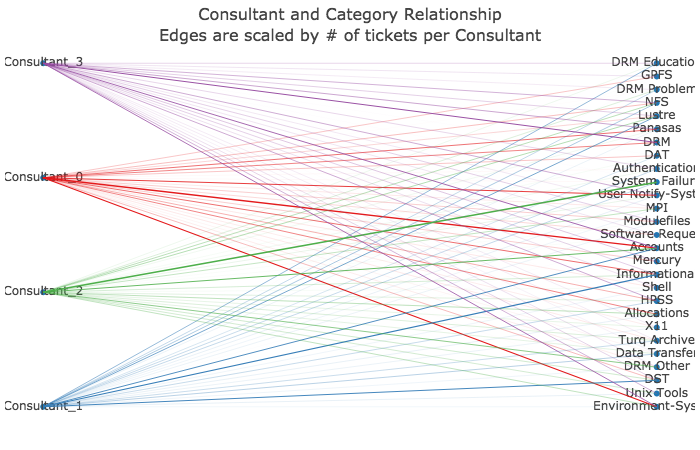}
	\caption{A graph showing the relationship between consultants and the categories from their tickets. For ease of reading we only used categories that had at least 200 tickets. The lines are scaled by consultant to accommodate for differences in the number of tickets (i.e. Consultant 1 might have more tickets than Consultant 2 assigned category X, but a higher percentage of Consultant 2's tickets deal with category X.)}
	\label{fig:cat_consult_graph}
\end{figure}

As shown in Figure \ref{fig:category_distribution}, most tickets are assigned to the broader categories such as ``Environment-System". The overall distribution of categories is interesting, but we were also curious about the distribution on a per-consultant basis, shown in Figure \ref{fig:cat_consult_graph}. As expected, the general categories are more popular than specific ones like ``Shell," and consultants tend to specialize and gravitate towards certain types of problems more than others. The preference of general categories over specific ones questions the need for specific categories at all. Next, we use text clustering algorithms to automatically generate categories.

\subsubsection{Defining New Categories}
Since the most used categories are vague, we explored using the techniques behind our feature sets from category prediction (see Section \ref{sec:category_prediction}) to define new categories to replace the current list of 93 categories (see Figure \ref{fig:category_distribution}). We used LDA topic modeling and Word2Vec to cluster words in the combined content of the entire dataset (i.e. subject, create message, content, and comments of all non-rejected tickets) and then used consultant input to add labels. Word2Vec is part of Doc2Vec, but is only the vector representations of words and not entire documents \cite{DBLP:journals/corr/MikolovSCCD13, Mikolov2013EfficientEO}. 

\paragraph{LDA Topic Modeling} In topic modeling, ``topics" are analogous to categories. For more human-readable topics, we set the topics at 10 and only used tokens with alphabetical characters (see Table \ref{tab:cleaning_regex}). The Consult Team helped assign labels to the words from each topic. Overall, we were very pleased with the generated categories, each one seemed coherent and did not require much imagination to assign a label. Example generated categories are in Table \ref{tab:cat_lda}.

\paragraph{Word2Vec} Word2Vec uses a neural network to create a vector representation of words based on their context (i.e. surrounding words). We used the Gensim implementation of Word2Vec \cite{rehurek_lrec}. Unlike the LDA categories, clustering the Word2Vec vectors did not return any useful or easily identifiable categories (Table \ref{tab:cat_word2vec}). We tried multiple clustering methods (KMeans, KMedoids, DBSCAN) and distances (Euclidean and cosine), and the KMedoids clustering of cosine distances returned the most coherent clusters. We also tried different ways of picking the words to represent each cluster. Using the word's frequency in the dataset produced more coherent clusters than using a word's distance from the cluster center or using a weighted combination of the two metrics. In addition to clustering the vectors, we found similar words using cosine similarity on the word's vector representation. The model recognized relationships in the data and was able to identify cluster names. Example queries are shown in Table \ref{tab:word2vec_query}. We were surprised that the model did not generate useful clusters, since it works very well on an individual word basis.

\begin{table*}
\centering
    \caption{A sample of LDA topic modeling auto-generated categories (topics=10, alpha=10, beta=1.0, iterations=2000, stopwords removed ). Note that some words are stemmed (e.g. \{module, modules\} $\rightarrow$ modul). Descriptions were added by consultants.}
    \begin{tabular}{cc}
    \toprule
    \textbf{Description} & \textbf{Keywords} \\ \midrule
    ASC Trinity and DATs & node trinity subject run dat test pm cc job reserv start request time knl trinitite \\ \midrule
    Group Accounts & account group user request project access hpc subject email approv pm rt machin submit add \\ \midrule
Filesystem and Storage & file unknown directori space error data hpss archiv ls permiss rank root info user scratch copi delet quota mon tri \\ \midrule
User Application & error modul load compil file code version mpi line work intel librari instal tri build status set problem number command \\ \midrule
Job Submission & job node run time user msub submit error script problem alloc moab state standard start queue slurm jobid process \\ \midrule
DST & system dst user schedul ant hpc work team inform comput complet time end offic avail action support \\ \midrule
Cluster login and Tri-lab access & ssh connect tri work access file network server user machin password rt problem system log port email \\
    \bottomrule
    \end{tabular}
    \label{tab:cat_lda}
\end{table*}

\begin{table}
    \caption{Auto-generated categories using Word2Vec with KMediods. The most frequent words were chosen to represent each cluster.}
    \begin{tabular}{cc} \toprule
    \textbf{Cluster} & \textbf{Keywords} \\ \midrule
   1 & node ing access system time trinity link wed ssh \\ \midrule
    2 & hpc machin thu cc default connect import script \\ \midrule
    3 & email project back ace turquois close yellow week \\ \midrule
    4 & \scale{0.35}{job file error unknown updat status start product submit network} \\ \midrule
    5 & run pm test code check instal limit trinitite today server \\ \midrule
    6 & problem dst schedul fix send phone day permiss dec morn \\ \midrule
    7 & \scale{0.35}{team modul info command comput mon fail process complet fri} \\ \midrule
    8 & user subject request directori issu set cielo creat tue data \\ \midrule
    9 & assign child custom number question make date hour \\ \midrule
    10 & work group rt account messag open load url end \\
    \bottomrule
    \end{tabular}
    \label{tab:cat_word2vec}
\end{table}

\begin{table}
\centering
    \caption{Example queries and results in our Word2Vec model. ``DST" is ``Dedicated System Time", ``red" is the name of the secure network at LANL and ``Grizzly" is one of the HPC machines. Words are similar if their cosine similarity is close to 1.}
    \begin{tabular}{cc}
    \toprule
    \textbf{Query} & \textbf{Similar Words} \\ \midrule
    dst & \scale{0.35}{outag dsts upgrad mainten downtim decommiss noon reschedul aam reboot} \\ \midrule
    red & \scale{0.35}{secur turquois yellow turqois registri turq turquos redston neural ibf} \\ \midrule
    grizzly & \scale{0.35}{snow wolf badger kodiak mustang moonlight woodchuck pinto trinitite trinity} \\
    \bottomrule
    \end{tabular}
    \label{tab:word2vec_query}
\end{table}

\section{Suggested Improvements}\label{sec:improvements}
Throughout the literature review process, ticketing tool development, and ticket exploration, we encountered situations where one additional data field would provide significantly more leverage, or where a pain point was an obvious candidate for streamlining. These improvements range from saving consultants a minute here and there to larger-scale projects that would require inter-team development.

\paragraph*{Access Outside User Information} Having access to information about the user, such as user's division and cluster allocations, would help narrow down a problem and make ticket creation quicker. In the consultant's current system, information about the user must be scavenged from the internal phone book, internal accounts and log resources, and from the user themself. Users can be vague and not provide enough detail in a message, and instead of wasting time with ``what machine are you running on?" it would be helpful if there was an integrated system that had the user's machine allocations and recent jobs to start from. This information would also make ticket creation quicker and allow consultants to focus on solving the problem, if they have access to user's information and do not have to ask more questions. Also, this information would be helpful for community analysis. Is there a high number of users from group X that need help? This information is useful because workshops could be geared towards those users, reducing their questions in the future.

\paragraph*{Access Outside System Information} Similar to the user information idea, it would be helpful to have immediate access to system states mentioned in a ticket. For example, an incoming ticket with the subject ``Snow lustre4 slow?" contains references to the cluster Snow and the filesystem lustre4. A suggestion of recent and current issues with those systems could save up to hours, as accessing that information requires asking system admins outside the team to check logs and internal dashboards. This information access would reduce the time it takes to solve issues for the users and the systems.

\paragraph*{Access Outside Vender/Software Information} Similar to accessing outside user and system information, it would be helpful to have information about recent updates or patches for software. Instead of spending time narrowing down an issue to user error or software error, seeing the most recent (or relevant) software patches or updates could take the guesswork out. This would contribute to finding a solution faster.

\paragraph*{Explicit QA Pairs} Similar to the problem with developing a tool for automatic ticket reply (Section \ref{sec:auto_reply}), we do not have a dataset for explicit question and answer pairs. Each ticket has a question from the user and a solution from the consultant, but the question is somewhere in the ``create message" (initial message from the user in emailed tickets). The problem with depending on a create message for a question is sometimes the user doesn't know the question to ask and the problem is vague. Also, we only have access to the create message for emailed tickets, since the first message in a phone call ticket would be a follow-up from the consultant, and might not include all the information discussed on the phone. A solution to lack of data would be to include a question and solution field for the consultant to fill out at the end of the ticket. We understand this would put more time per ticket, but the data collected would pay off later. It would take a few minutes to enter the information, but it would reduce the solution time on future issues in the form of a tool that can suggest solutions based on the user's problem and in the form of information to add to documentation, further reducing the need for user questions in the first place.

\paragraph*{Match the Business Logic to Best Practices} The ticketing system should reflect the practices of the consultants. For example, in order to have accurate reports, the desired information from the tickets (e.g. Category) should be required by the system before a ticket is marked as ``resolved." This would help with consistency for reporting and accuracy, as even though information could be added later on, it would be most accurate when the ticket is fresh in the consultant's mind.

\paragraph*{Procedure for Template Use} This point is more of an open discussion on the use of templates than a suggestion. As discussed in Section \ref{sec:auto_reply}, we were not able to create a model to predict which template should be sent to an incoming message because our ticket dataset did not contain enough templates. After discussing this with the Consult Team, we were told that the problems that consultants receive have such a variety that there usually isn't a standard solution, and there are only a small handful of common problems that already have a form-based process (e.g. opening a port). 

The basic concept of a template is to save a consultant time from typing out a standard reply. In the ticketing systems, the reply is usually a suggested solution. We think a different approach is to use a template to help clarify a user's problem. Instead of offering a solution, a template could be selected based on the user's incoming message to ask for supplementary information. Very often a user will ask about a job or software installation but does not immediately provide pertinent information such as which cluster they are working on, which software version they need, etc. Above we suggested a more integrated system with user and system information, but an immediate resolution could be a consultant-curated template that is sent to the user to ask for that information. This would allow the knowledgeable consultant to ask for the information instead of allowing an unsure user to fill out an incorrect form before talking to anyone.

A template could also be used to simply point to relevant documentation to send to the user while the consultant collects more information. If the information on the pages solves their problem, then the ticket can be closed and the consultant doesn't have to waste time on an already ``solved" problem. And if the problem isn't answered by the information in the links, then the consultant has feedback on how to improve the pages. This self-help system is not a novel idea, and is implemented by many online customer support tools. These reference systems help consolidate information, and can help with the onboarding of future consultants.

Also, these templates should be searchable and possibly categorizable.  A template does not save a consultant time if they have to scroll through a large list of templates to use. Once a dataset of incoming message / template pairs is created, a tool can be created to suggest a template, making template selection even quicker, hence solving a user's problem quicker.

\paragraph*{Assign Ticket Category at End of Ticket} As discussed in Section \ref{sec:category_prediction}, we developed a model to assign a category to a ticket based on its create message and subject line. The model did not do very well, and as there could be room for improvement in the development and training of the model, we also think how tickets are assigned categories could be improved, as well as the categories themselves. Certain categories are vastly more popular than others, and that relates to the vagueness of the category (e.g. ``environment" vs ``MPI"). Category vagueness can lead to overlap, which provides multiple valid options for a ticket. We think a policy of assigning a category when closing a ticket would be more helpful than assigning one earlier, because then the consultant knows exactly what the ticket entailed and already narrowed down the user's problem. This would help with accuracy for reporting and does not add any time, as it only moves the upfront cost of adding a Category to the end.

\paragraph*{Assign Multiple Categories to a Ticket} The idea of how categories should be used came up in our discussions. Since they are useful for reporting, then giving someone the option to assign a ticket to multiple categories could be useful. A problem can cover multiple areas, and choosing only one forces the categorizer to be vague or spend more time than they want. This ties into the category suggestion model in Section \ref{sec:category_prediction} since one could just select multiple suggestions. Using multiple categories would provide more detailed reporting, shedding more light on types of user issues.

\paragraph*{Advanced Ticket Search} The ability to search past tickets would be helpful when a consultant knows a problem has been encountered before. Ideally, a ``similar tickets" feature would help with this, but searching through tickets using advanced queries like those available in Google or Elasticsearch would save time in finding relevant information that could be used in solving the current ticket more quickly.

\section{Conclusion}
HPC support teams are the first line of defense against large-scale problems; supporting these teams with integrated systems and an intuitive request tracking system is critical. In our work we presented tools to help automate some of the ticketing workflow, explored tickets to find community trends and define categories, and discussed requirements and desirables for future ticketing systems.

For automating ticket workflow we tested machine learning-based  text analysis techniques for automatic category assignment of an incoming ticket (top performance of 72\% ``accuracy@3", see Section \ref{sec:category_prediction}), applied those same text analysis techniques for recommending similar tickets for quicker problem-solving (results in Section \ref{sec:ticket_recommend}), and attempted to create a tool to suggest a template reply based on the incoming ticket (see issues in Section \ref{sec:auto_reply}). Our tools are not yet deployed, but could be developed into plug-ins for a new ticketing system. In our ticket exploration, we applied the same machine learning techniques from category prediction to define a new set of categories for reporting. The requirements and desirables for a new ticketing system include a discussion of features whose impact would save user support teams time with researching user problems as well as suggestions for tracking ticket information and solutions for reporting and trend analysis.

We hope our work and recommendations inspires more ideas and collaborative tool-building, as well as encouraging other institutions to publicly release details of their ticketing systems and analyses.

\section*{Acknowledgments}
The authors thank Mike Mason, Ed Rose, Malachy O'Connell, and Hugh Greenberg for access to data and the NERSC ticketing team for loose conceptual collaboration. We also thank Mike Coyne and Rob Derrick for letting Alexandra shadow them, and Ben Santos, Lena Lopatina, and Hunter Easterday for their suggestions and time reviewing results. 

\bibliographystyle{IEEEtran}
\bibliography{IEEEabrv,references}

\end{document}